\documentclass[10pt]{revtex4}
\usepackage{graphics}
\usepackage{graphicx}
\usepackage{dcolumn}
\usepackage{bm}

\newcommand{\de}{\partial}
\newcommand{\sech}[1]{\textrm{sech}\left(  #1\right)}
\newcommand{\eq}[2]{\begin{equation} \label{#1} #2 \end{equation}}

\newcommand{\etal}{{\em et al.}}

\newcommand{\eee}{\mathbf{e}}
\newcommand{\fff}{\mathbf{f}}

\newcommand{\rr}{\mathbf{r}}
\newcommand{\rrort}{\mathbf{r}_{\bot}}

\begin{document}

\title{Emergence of geometrical optical nonlinearities in photonic crystal fiber nanowires}

\author{Fabio Biancalana, Truong X. Tran, Sebastian Stark, Markus A. Schmidt and Philip St. J. Russell}
\affiliation{Max Planck Institute for the Science of Light, G\"{u}nther-Scharowsky Str. 1, Bau 26, 91058 Erlangen, Germany}

\begin{abstract}
We demonstrate analytically and numerically that a subwavelength-core dielectric photonic nanowire embedded in a properly designed photonic crystal fiber cladding shows evidence of a previously unknown kind of nonlinearity (the magnitude of which is strongly dependent on the waveguide parameters) which acts on solitons so as to considerably reduce their Raman self-frequency shift. An explanation of the phenomenon in terms of indirect pulse negative chirping and broadening is given by using the moment method. Our conclusions are supported by detailed numerical simulations.
\end{abstract}

\maketitle

Photonic nanowires (PhNs), i.e. dielectric waveguides with a sub-wavelength core diameter, tight mode confinement and strong
waveguide dispersion, have recently attracted a growing interest due to accessibility of new fabrication techniques for a large variety of materials, which may lead to a number of miniaturized, high-performance photonic devices \cite{nanowires}. The small effective modal area exhibited by PhNs, which increases considerably the Kerr nonlinear coefficient, and the degree of controllability of the dispersion characteristics, make PhNs especially suitable for the investigation of extreme nonlinear phenomena such as supercontinuum generation, as many optical solitons can be excited by using small pump energies \cite{nonlinearnanowires}.

In recent theoretical work, a novel propagation equation that accurately describes the nonlinear evolution of light pulses in PhNs was introduced \cite{truong}, see also Eq. (\ref{finalx1}) of the present Letter. The fundamental feature found in \cite{truong} is that, thanks to the fact that the correct equation takes into full account of variations in the linear mode profiles of the waveguide with wavelength, new nonlinear effects arise in PhNs, unknown in previous formulations based on the generalized nonlinear Schr\"odinger equation (GNLSE) \cite{previous}, which all assume fixed transverse field profiles. Most of the additional terms described in Ref. \cite{truong} have been found to have an extremely small magnitude, so that they can be safely neglected for large core fibers. However, the longitudinal component of the electric field of the fundamental mode of PhNs becomes progressively more important when decreasing the core diameter or when increasing the refractive index contrast between core and cladding \cite{truong}. This vigorously breaks the rotational symmetry of the mode, whose transverse profile becomes very sensitive to frequency, thus making the new {\em geometrical nonlinearity} - as we shall call it in the following - extremely important. In fact, for small enough core sizes, such new nonlinear terms may even enter into strong competition with the Raman effect term for some range of frequencies.

Only the circular geometry for PhNs has been considered in Ref. \cite{truong}. Circular strands of high-refractive index materials in air or in a homogeneous cladding, however, are far from optimal for detecting experimentally the effects of the novel term, because the maximum of the geometrical nonlinear coefficient is located, as a rule, far away in a region of strong normal dispersion, where bright solitons cannot exist. In fact, as has been anticipated in Ref. \cite{truong}, the new nonlinearity is visible only in presence of solitons, and its effects being nearly invisible in the normal dispersion regime.

\begin{figure}[htb]
\centerline{\includegraphics[width=8cm]{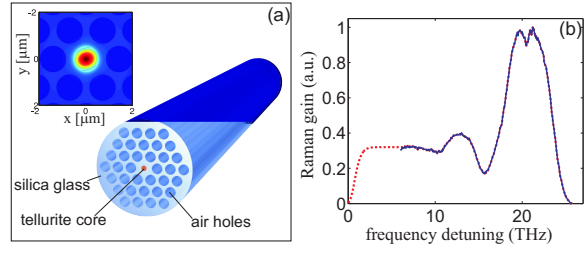}}
\caption{\small (Color online) (a) Schematics of the proposed PCF nanowire geometry. The central nanowire core is made of tellurite glass ($77$TeO$_{2}$-$10$Na$_{2}$O-$10$ZnO-$3$PbO composition, T2 glass from Ref. \cite{ramantellurite}), while the cladding is made of silica with a triangular lattice of air holes. Parameters are: pitch $\Lambda=1.4$ $\mu$m, hole radius $R=0.56$ $\mu$m, core radius $R_{c}=0.5$ $\mu$m. Inset, the mode profile $||\hat{\eee}(\rr_{\perp})||$ of the waveguide for $\lambda=1.55$ $\mu$m is shown.
(b) Profile of the Raman gain spectrum for the T2 composition. Solid blue line: experimental data taken from Ref. \cite{ramantellurite}. Dotted red line: fit used in our simulations.} \label{fig1}
\end{figure}

However, one is by no means restricted to the use of a homogeneous cladding around the high-index core. Here we explore the possibility of introducing the PhN into a silica-based photonic crystal fiber (PCF, \cite{russell}) with a triangular arrangement of the holes [see Fig. \ref{fig1}(a)]. We call such a design a {\em photonic crystal fiber nanowire} (PCF-NW). Such a structure has the additional advantage that relatively long PhNs can be supported by the robust PCF cladding \cite{russell,wolchover}. In this Letter we theoretically demonstrate that, by means of careful choice of parameters in the design of PCF-NWs, one can move the maximum of the geometrical nonlinearity {\em inside} the region of anomalous dispersion. This will make it much easier to observe the new geometrical nonlinearity experimentally, in that it causes considerable suppression of the soliton self-frequency shift (SSFS, see Ref. \cite{gordon}) from the very start of the propagation. Thus, just a judicious choice of the geometry of the PCF cladding around the nanowire, which affects the dispersive properties of the waveguide and the way the mode profiles change with wavelength, is sufficient for the emergence of this completely unexplored optical nonlinearity, a feat that would be extremely difficult if not impossible to achieve with other simpler geometries.

The fiber geometry that we propose in this Letter is shown in Fig. \ref{fig1}(a). It is made of a silica PCF cladding with a triangular lattice of air holes, with pitch $\Lambda=1.4$ $\mu$m and hole radius $R=0.56$ $\mu$m. The central core, of radius $R_{c}=0.5$ $\mu$m, is made of a high refractive index tellurite glass (T2 composition taken from Ref. \cite{ramantellurite}), which possesses an estimated nonlinear coefficient $n_{2}\sim 4\times 10^{-19}$ m$^{-1}$W$^{-1}$, almost $15$ times larger than that of fused silica \cite{telluritebook}. Its Raman gain spectrum $h(\omega)$ is shown in Fig. \ref{fig1}(b), and has a large peak centered around $20$ THz. In this waveguide, light confinement is provided by total-internal-reflection at the core-cladding boundary. Air holes in the rings modify the dispersion in such a way as to match our requirements, as will be explained shortly. Typically, only one ring of holes is sufficient to obtain the dispersive features described below.
Our choice of core material and PCF parameters has been dictated by two conditions, which cannot easily be simultaneously met. The first requirement is that there must be a relatively large refractive index contrast between core and cladding, so that the magnitude of the longitudinal component of the electric field becomes appreciable, and in turn that the mode profile changes strongly with frequency. The second requirement is that the holes should considerably modify the GVD, in such a way that the maximum of the geometrical nonlinear coefficient is located inside the region of anomalous dispersion. However, the larger the refractive index contrast, the more the field is localized in the core, which makes the dispersion of the waveguide progressively more and more like that of a single rod surrounded by homogeneous silica, which has been proved to be non-optimal in Ref. \cite{truong}. A trade-off between the above conditions must be found. We have examined the dispersion of many high-index glasses for the core material (chalcogenide, bismuth, germanium-doped silica), and even some nonlinear liquids such as carbon disulfide, and systematically explored hundreds of specific parameters for the PCF cladding, but the above tellurite-based design seems to be one of the best solutions for our purposes.
Such multi-glass hybrid waveguides can be fabricated by using the pressure-cell approach \cite{telluriteinfibers}. Melted tellurite glass is pressed under large external gas pressure into the holes of silica PCFs. The technique relies on the fact that the tellurite glass has a significantly lower softening point than silica, leaving the geometry defined by the silica host unchanged during filling \cite{fillingtechnique}. The problem of surface-induced glass crystallization in tellurite compound glasses has been solved in recent experiments and will be published elsewhere. Altogether, the pressure-cell technique provides a potential way to realize the proposed structures.

For each frequency $\omega$, the linear fundamental mode of the waveguide of Fig. \ref{fig1}(a) has a normalized electric field profile given by $\hat{\eee}_{\omega}(\rr_{\perp})$, where $\rr_{\perp}$ are the transverse coordinates. The inset in Fig. \ref{fig1}(a) shows a contour plot of the norm $||\hat{\eee}||\equiv[\hat{e}_{x}^2+\hat{e}_{y}^2+\hat{e}_{z}^2]^{1/2}$ for the fiber parameters given in the caption, corresponding to our representative PCF-NW design that we shall use throughout the paper. Due to the invariance of the PCF cladding under the $C_{6v}$ symmetry group, the waveguide does not exhibit any birefringence, and it has a fundamental mode that is degenerate in the two orthogonal polarization states. The crucial point of our formalism is that one can describe the full $\omega$-variations of $\hat{\eee}$ through the Taylor series
\eq{expansion1}{\hat{\eee}_{\omega}(\rr_{\perp})=\sum_{j\geq
0}\frac{1}{j!}\fff^{(j)}_{\omega_{0}}(\rr_{\perp})\left(\frac{\Delta\omega}{\omega_{0}}
\right)^{j}} where $\Delta\omega\equiv\omega-\omega_{0}$ is the
frequency detuning from an arbitrary reference frequency $\omega_{0}$, and the quantity $\fff^{(j)}_{\omega_{0}}\equiv\left[\omega_{0}^{j}\de^{j}\hat{\eee}_{\omega}(\rr_{\perp})/\de\omega^{j}\right]_{\omega=\omega_{0}}$ is proportional to the $j$-th frequency derivative of
the mode profile. From now on, letters $jhpv$ will be used for derivative indices. Following Ref. \cite{truong}, one can rigorously prove that the equation governing the nonlinear light propagation of one of the two polarization states of the fundamental mode of the PCF-NW is given by
\eq{finalx1}{i\de_{z}Q+\hat{D}(i\de_{t})Q+\sum_{jhpv}\gamma^{jhpv}
\hat{G}^{j}(i\de_{t})\phi^{hpv}=0.} In Eq. (\ref{finalx1}), $Q(z,t)$ is the electric field envelope, $\hat{D}(i\de_{t})\equiv\beta(\omega_{0}+i\de_{t})-\beta(\omega_{0})$
is the dispersion operator that encodes all information on the
fiber GVD around $\omega_{0}$ \cite{agrawalbook}, $\beta$ is the $\omega$-dependent propagation constant of the fundamental mode, and
$\hat{G}^{j}(i\de_{t})\equiv\left[1+(i/\omega_{0})\de_{t}\right]
\left[(i/\omega_{0})\de_{t}\right]^{j}$
is an operator that naturally contains the dynamics of the shock term at the zero-th order of the Taylor expansion ($j=0$). The {\em convoluted nonlinear fields} used in Eq. (\ref{finalx1}) are defined as:
\eq{finalx4}{\phi^{hpv}(z,t)\equiv\frac{\left[(i\de_{t})^{h}Q\right]\left\{R\otimes\left(\left[(i\de_{t})^{p}Q\right]\left[(-i\de_{t})^{v}Q^{*}\right]\right)\right\}}{\omega_{0}^{h+p+v}},}  where symbol $\otimes$ is used to denote a time-convolution: $A\otimes B\equiv\int_{-\infty}^{+\infty}A(t-t')B(t')dt'$=$B\otimes A$.
In Eq. (\ref{finalx4}), $R(t)\equiv(1-\theta)\delta(t)+\theta h(t)$ is the nonlinear response function of the core, which is made of a tellurite glass following the design of Fig. \ref{fig1}(a), and includes the instantaneous Kerr [proportional to the Dirac delta $\delta(t)$] and the non-instantaneous Raman [proportional to $h(t)$] responses exhibited by the core material, $\theta$ being the relative importance between the two. In this Letter, from the experimental Raman gain of tellurite glass (solid blue line in Fig. \ref{fig1}(b), taken from Ref. \cite{ramantellurite}) we can extract a fit of $h(\omega)$ [dotted red line in Fig. \ref{fig1}(b)] that we thus use in the numerical simulations. The $\rr_\perp$ dependence of $R(t)$ can be safely neglected, since most of the energy is located in the core material, which is also much more nonlinear than the surrounding silica cladding. Note that in expression (\ref{finalx4}), $\phi^{000}=Q\int R(t-t')|Q(t')|^{2}dt'$ gives the conventional, zero-th order convolution that is used in the GNLSE \cite{gaeta}.

\begin{figure}[htb]
\centerline{\includegraphics[width=8.5cm]{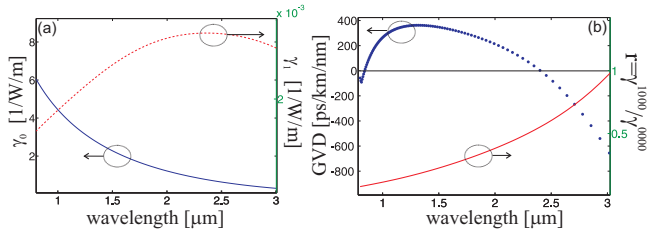}}
\caption{\small (Color online) Linear and nonlinear data for the PCF-NW design of Fig. \ref{fig1}(a). (a) Blue solid and red dashed lines indicate respectively $\gamma_{0}$ and $\gamma_{1}$ versus wavelength. (b) Blue dots indicate the GVD of the waveguide, with zeros-GVD points located at $\lambda\simeq 0.84$ $\mu$m and $\lambda\simeq 2.4$ $\mu$m. Red solid line indicates parameter $r=\gamma^{1000}/\gamma^{0000}$, that appears in Eq. (\ref{qq2}).} \label{fig2}
\end{figure}

The last ingredient in Eq. (\ref{finalx1}) contains the generalized nonlinear coefficients $\gamma^{jhpv}$, defined as
\begin{widetext}
\begin{eqnarray}
\gamma^{jhpv}(\omega_{0})\equiv\frac{\omega_{0}}{16c}\int
d\rr_{\perp}\chi^{(3)}_{xxxx}(\rrort)\frac{\left\{[\fff^{*(j)}_{\omega_{0}}\cdot\fff^{(h)}_{\omega_{0}}
]
[\fff^{(p)}_{\omega_{0}}\cdot\fff^{*(v)}_{\omega_{0}}]+[\fff^{*(j)}_{\omega_{0}}\cdot\fff^{(p)}_{\omega_{0}}
]
[\fff^{(h)}_{\omega_{0}}\cdot\fff^{*(v)}_{\omega_{0}}]+[\fff^{*(j)}_{\omega_{0}}\cdot\fff^{*(v)}_{\omega_{0}}
]
[\fff^{(h)}_{\omega_{0}}\cdot\fff^{(p)}_{\omega_{0}}]\right\}}{j!h!p!v!},\label{gammacapitalx2}
\end{eqnarray}
\end{widetext}
where $\chi^{(3)}_{xxxx}(\rr_{\perp})$ is the third-order susceptibility, which is a function of the transverse coordinates in the waveguide of Fig. \ref{fig1}(a). Definition (\ref{gammacapitalx2}) is a generalization of the nonlinear coefficient commonly used in fiber optics \cite{agrawalbook}, and takes into account the full vector nature of the field profile as well as its frequency variations. Such variations are at the very core of the new geometrical nonlinearities described here, since the Taylor series of Eq. (\ref{expansion1}) implies the existence of an infinite number of additional nonlinear terms that depend on progressively higher-order time derivatives of the envelope. The quantities $\gamma^{jhpv}$ satisfy general symmetry relations  that drastically reduce the number of independent nonlinear coefficients for each order of the derivative \cite{poletti,truong}: $\gamma^{jhpv}=\gamma^{vhpj}=\gamma^{jphv}=\gamma^{vphj}$, $\gamma^{jhhh}=\gamma^{hjhh}$, $\gamma^{jhhj}=\gamma^{hjjh}$. In Eqs. (\ref{expansion1}) and (\ref{finalx4}), each field derivative is associated with a factor $\omega_{0}^{-1}$, which ensures convergence. Thus the physically meaningful nonlinear coefficients can be defined as $\tilde{\gamma}^{jhpv}\equiv\gamma^{jhpv}(\omega_{0}t_{0})^{-(j+h+p+v)}$, where $t_{0}$ is the input pulse duration.

From the expression in Eq. (\ref{gammacapitalx2}) one can define the zero-th order nonlinear coefficient of the waveguide $\gamma_{0}\equiv\gamma^{0000}=\tilde{\gamma}^{0000}$, corresponding to the conventional definition used in nonlinear fiber optics \cite{agrawalbook}. To first order in the Taylor expansion in Eq. (\ref{finalx1}) one can define the coefficient $\gamma_{1}\equiv\tilde{\gamma}^{1000}=\tilde{\gamma}^{0100}=\tilde{\gamma}^{0010}=\tilde{\gamma}^{0001}=\gamma^{1000}/(\omega_{0}t_{0})$, associated with nonlinear convoluted fields $\phi^{hpv}$ that contain only one time derivative of the envelope. Fig. \ref{fig2}(a) shows plots of $\gamma_{0}$ and $\gamma_{1}$ versus reference wavelength for the fiber design of Fig. \ref{fig1}(a). The fiber GVD is shown in Fig. \ref{fig2}(b) with blue dots. It is clear from this figure the well-known fact that $\gamma_{0}$ decreases monotonically for longer wavelengths \cite{dudleyarea}. However, it is interesting to note that the geometrical nonlinear coefficient $\gamma_{1}$ initially increases, but then reaches a maximum near the infrared zero-GVD point of the fiber (located at $\lambda\simeq 2.4$ $\mu$m), and then tends to zero for even longer wavelengths. The close vicinity of $\max(\gamma_{1})$ to the anomalous GVD of the fiber is an atypical feature, that we have found only in a few very specific designs, including the one presented in Fig. \ref{fig1}(a). The `normal' situation, which is also true for circular PhNs surrounded by homogeneous media (such as, for instance, tapered fibers), is that $\max(\gamma_{1})$ is located well within the region of normal GVD \cite{truong}.

\begin{figure}[htb]
\centerline{\includegraphics[width=9cm]{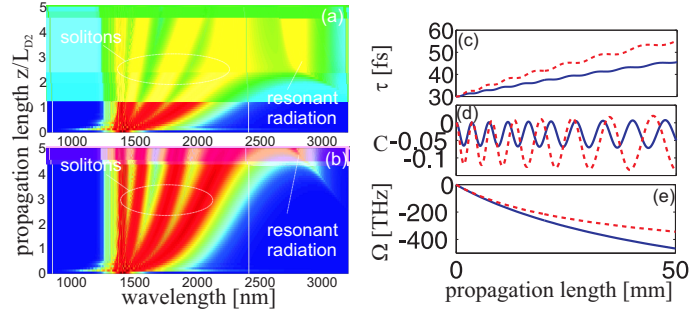}}
\caption{\small (Color online) (a) Evolution of a $t_{0}=150$ fs pulse in the waveguide of Fig. \ref{fig1}(a) according to Eq. (\ref{finalx1}) truncated at the $0$-th order, i.e. by using the conventional GNLSE. Second order dispersion length is $L_{D2}\simeq 6$ cm. (b) Same as (a) but truncating Eq. (\ref{finalx1}) at the $2$-nd order. The suppression of SSFS in the latter case is evident. Vertical white lines indicate the two zero-GVD wavlengths. (c,d,e) $z$-evolution of pulse duration $\tau$ [$t_{0}=\tau(0)=30$ fs], chirp $C$ and soliton frequency shift $\Omega$ according to Eqs. (\ref{qq1}-\ref{qq3}), for a $N=7$ soliton, pump wavelength $\lambda=1.4$ $\mu$m. Blue solid (red dashed) line refer to the case $r=0$ ($r=0.2$) in Eqs. (\ref{qq1}-\ref{qq3}).} \label{fig3}
\end{figure}

We now show that the onset of geometrical nonlinearities leads to a strong suppression of the SSFS. In order to do this, we compare two direct numerical simulations, the first one obtained by truncating the Taylor expansion of Eq. (\ref{finalx1}) to the zero-th order (1 nonlinear convolution, corresponding to the conventional GNLSE) [Fig. \ref{fig3}(a)], the second one obtained by truncating the same sum to the second order (15 nonlinear convolutions in total), which thus takes into account the dominant terms of the geometrical nonlinearities [Fig. \ref{fig3}(b)]. It is easily seen by comparing Fig. \ref{fig3}(a) with Fig. \ref{fig3}(b) that the net effect of the additional nonlinearities is to reduce considerably the SSFS in the fiber. Thus a unique interplay between the geometrical nonlinearities and the SSFS takes place in properly designed PCF-NWs.

It is possible to qualitatively understand the reason of the above SSFS suppression mechanism by using the so-called moment method \cite{agrawalbook,moment}. One assumes that after formation, each solitonic pulse does not change its functional shape, given by $Q(t)=[Pt_{0}/\tau]^{1/2}\sech{[t-T]/\tau}e^{-i\Omega(t-T)-iC(t-T)^{2}/(2\tau^{2})}$, where $T$ is the temporal delay of the solitonic pulse, $\Omega$ is its frequency detuning from the reference frequency $\omega_{0}$, $\tau$ is the soliton pulse width, $C$ is the pulse chirp, $\beta_{2}=(\de^{2}\beta/\de\omega)_{\omega=\omega_{0}}<0$ is the second-order (anomalous) dispersion coefficient, $P\equiv N^{2}P_{0}$ is the peak power, $P_{0}\equiv|\beta_{2}|(t_{0}^{2}\gamma_{0})^{-1}$ is the fundamental ($N=1$) soliton power, $N$ is the soliton order and $T_{R}\equiv\int_{-\infty}^{+\infty}R(t)dt$ is the first moment of the Raman response. One can prove that the $z$-evolution of $\tau$, $C$ and $\Omega$ is then given by the following closed set of equations:
\begin{eqnarray}
\frac{d\tau}{dz}&=&\beta_{2}\frac{C}{\tau}, \label{qq1} \\
\frac{dC}{dz}&=&\frac{4|\beta_{2}|N^{2}}{\pi^{2}t_{0}\tau}\left[\left(1-\frac{t_{0}}{\tau}\right)+\frac{\Omega}{\omega_{0}}(1+4r)\right]\label{qq2}\\
\frac{d\Omega}{dz}&=&-\frac{8T_{R}N^{2}|\beta_{2}|}{15t_{0}}\frac{1}{\tau^{3}},\label{qq3}
\end{eqnarray} with the initial conditions $\tau(0)=t_{0}$, $C(0)=\Omega(0)=0$. Higher-order terms in the dispersion and small terms proportional to $C^{2}$ have also been neglected for sake of clarity. Geometrical nonlinearities are parameterized by the coefficient $r\equiv\gamma^{1000}/\gamma^{0000}$, the only one that appears in Eq. (\ref{qq2}), shown in Fig. \ref{fig2}(b). Note that the condition $r>1/4$ is necessary for the new nonlinearity to dominate the shock term in Eq. (\ref{qq2}).
Equations (\ref{qq1}-\ref{qq3}) are written under the simplifying assumption that the GVD does not change during the soliton evolution, and that the pulse duration is longer than $100$ fs, so that one can use a well-known approximate expression for the Raman term \cite{agrawalbook}. The term on the right-hand side of Eq. (\ref{qq3}), proportional to $T_{R}$, is responsible for the constant SSFS along the fiber \cite{gordon,moment}. The rate of this shift is very sensitive to the actual pulse width, since it is determined by $\tau^{-3}$, and it is always directed towards negative detunings, i.e. towards the red part of the spectrum. However, due to the action of the right-hand side of Eq. (\ref{qq2}), the soliton acquires a small chirp even if its initial chirp vanishes. The slope of this chirp is initially negative, due to the initial condition $\tau(0)=t_{0}$, which makes the term proportional to $\Omega/\omega_{0}$ dominant. Thus $C<0$ in the initial stage of propagation, which in turn leads to a pulse broadening due to Eq. (\ref{qq1}), for which $\beta_{2}C>0$. Finally, such broadening leads to a sharp suppression of the SSFS given by Eq. (\ref{qq3}), due to the $\tau^{-3}$ dependence of its right-hand side. This mechanism qualitatively explains the overall suppression of the SSFS due to the geometrical nonlinearity on soliton propagation observed in the direct numerical simulations of Eq. (\ref{finalx1}). The long term behavior of the propagation can only be understood by numerically solving Eqs. (\ref{qq1}-\ref{qq3}). In Figs. \ref{fig3}(c-e) we show the $z$-evolution of the quantities $\tau$, $C$ and $\Omega$ for parameters given in the caption. $C$ undergoes several deep oscillations, and its sign is mostly negative throughout the whole propagation [Fig. \ref{fig3}(d)]. The pulse duration $\tau$ undergoes similar oscillations (with a smaller magnitude), but overall it constantly grows [Fig. \ref{fig3}(c)]. $\Omega$, however, is not too sensitive to such oscillations, due to the fact that its derivative never changes sign in Eq. (\ref{qq3}) [Fig. \ref{fig3}(e)].

In conclusion, we have shown the emergence of a new type of nonlinearity in tellurite PhN embedded in a PCF cladding, which strongly depends on the geometrical parameters of the PCF design and the specific dispersion of the core material. Apart from its value as a new fundamental entity in nonlinear fiber optics, the existence of such nonlinearity shows that there is still much unexpected physics to unveil in complex PCF geometries.

This work is supported by the German Max Planck Society for the Advancement of Science (MPG).

\end{document}